\documentclass[11pt]{article}

\usepackage[margin=1in]{geometry}
\usepackage{hyperref}
\usepackage[numbers]{natbib}
\usepackage{bm}
\usepackage{amsmath}
\usepackage{amsfonts}
\usepackage{amssymb}
\usepackage{booktabs}
\usepackage{multirow}
\usepackage{array}
\usepackage[justification=centering]{caption}
\usepackage{mathtools}
\usepackage{graphicx}
\graphicspath{{Figures/}}
\usepackage{float}
\usepackage{adjustbox}
\usepackage{verbatim}
\usepackage{siunitx,etoolbox}
\usepackage{rotating}
\usepackage{afterpage}
\usepackage{chngpage}

\newcommand{\EX}{\mathbb{E}}
\newcommand{\bbeta}{\bm{\beta}}
\def\*#1{\mathbf{#1}}
\newcommand{\RN}[1]{%
  \textup{\uppercase\expandafter{\romannumeral#1}}%
}
\newcommand*{\tran}{{^\prime}}

\DeclarePairedDelimiter\abs{\lvert}{\rvert}%
\DeclarePairedDelimiter\norm{\lVert}{\rVert}%
\makeatletter
\let\oldabs\abs
\def\abs{\@ifstar{\oldabs}{\oldabs*}}
\let\oldnorm\norm
\def\norm{\@ifstar{\oldnorm}{\oldnorm*}}
\makeatother
\DeclareMathOperator*{\argmin}{arg\,min}

\title{A Minimum Message Length Criterion for Robust Linear Regression}

\author{Chi Kuen Wong, Enes Makalic, Daniel F. Schmidt}

\begin{document}

\maketitle

\begin{abstract}
This paper applies the minimum message length principle to inference of linear regression models with Student-$t$ errors. A new criterion for variable selection and parameter estimation in Student-$t$ regression is proposed. By exploiting properties of the regression model, we derive a suitable non-informative proper uniform prior distribution for the regression coefficients that leads to a simple and easy-to-apply criterion. Our proposed criterion does not require specification of hyperparameters and is invariant under both full rank transformations of the design matrix and linear transformations of the outcomes. We compare the proposed criterion with several standard model selection criteria, such as the Akaike information criterion and the Bayesian information criterion, on simulations and real data with promising results.
\end{abstract}

\section{Introduction}
Consider a vector of $n$ observations $\*y=(y_1,\dots,y_n)\tran$ generated by the linear regression model
\begin{equation}
	\label{eq:tlinear}
	\*y = \beta_0 \*1_n + \*X \bbeta + \bm{\varepsilon} \,,
\end{equation}
where $\*X = (\*x_1,\dots,\*x_k)$ is the design matrix of $p$ explanatory variables with each $\*x_i \in \mathbb{R}^n$, $\bbeta \in \mathbb{R}^{p}$ is the vector of regression coefficients, $\beta_0 \in \mathbb{R}$ is the intercept parameter, $\*1_n$ is a vector of $1$s of length $n$, and $\bm{\varepsilon} = (\varepsilon_1,\dots,\varepsilon_n)\tran$ is a vector of random errors. In this paper, the random disturbances $\varepsilon_i$ are assumed to be independently and identically distributed as per a Student-$t$ distribution with location zero, scale $\tau$, and $\nu$ degrees of freedom. The Student-$t$ linear regression model finds frequent application in modelling heavy-tailed data~\cite{fernandez1998bayesian,jacquier2002bayesian}. Compared with the more popular Gaussian linear regression model the Student-$t$ linear regression model has been shown to have better performance in terms of reducing the negative effects caused by outliers~\cite{west1984outlier,lange1989robust}. 

This paper investigates parameter estimation and variable selection for the Student-$t$ linear regression model using the minimum message length (MML) principle~\cite{wallace2005statistical} and is an extension of the MML$_u$ model selection criterion described in~\cite{schmidt2009mml}. The MML$_u$ criterion is known to be closely related to Rissanen's normalized maximum likelihood (NML) denoising approach~\cite{rissanen2000mdl} and shares a number of its attractive properties (e.g., consistency)~\cite{schmidt2012consistency}. These two criteria are virtually indistinguishable when the signal-to-noise ratio is moderate to large. In contrast, when the signal-to-noise ratio is small, the NML-based approaches are known to perform poorly; due subtle differences between the complexity penalties of MML$_u$ and NML, the MML$_u$ approach is less prone to overfitting in these regimes. 

While the MML$_u$ criterion has been shown to perform well in a range of experiments using real and simulated data, it is limited to Gaussian linear regression models and is not invariant to translations of the target $\*y$. In this manuscript, we generalize the MML$_u$ approach to the Student-$t$ linear regression model and derive a new model selection criterion that is invariant to linear transformations of both the explanatory variables and the targets. The latter invariance property ensures that the criterion does not depend on the particular choice of the measurement units in which the data is recorded. 

\section{Minimum Message Length}
\label{sec:mml}
The minimum message length (MML) principle~\cite{wallace2005statistical,wallace1968information,wallace1975invariant,wallace1987estimation} was introduced by C. S. Wallace and D. M. Boulton in the late 1960s. The underlying philosophy of the MML principle is that one can use data compression to learn structure from data. It is well known that if there is little structure (regularities) in the data, the probability of greatly compressing the data is vanishing small (no hypercompression inequality~\cite{barron1985logically}). This result also implies that models which attain high levels of compression are likely to have captured real attributes of the underlying data generating mechanism.

Imagine one wishes to encode the observed data into a message string (e.g., a binary string) that is uniquely decodable. In the MML framework, this message string is called an explanation and consists of two components, namely the \emph{assertion} and the \emph{detail}. The assertion encodes a statement describing a statistical model (i.e., model structure and parameters) and the detail encodes the data using the model specified in the assertion. Inference within the MML framework proceeds by finding the model that results in the greater compression of the data; i.e., has the shortest two-part message. While computing the exact MML codelength for a given problem is in general NP-hard~\cite{farr2002complexity}, there exist a range of computationally tractable approximations for computing message lengths.



The most commonly used message length formula is the MML87 approximation~\citep{wallace1987estimation}. Suppose we are given a vector of data $\*y = (y_1,\dots,y_n)\tran$. Let $f(\cdot|\bm{\theta})$ denote the set of probability density functions indexed by the $k$-dimensional parameter vector $\bm{\theta} \in \Theta$ where $\Theta \subset \mathbb{R}^k$ is the parameter space. The message assertion needs to name a specific member from the uncountable set $\Theta$, however, from Shannon's theory information, codelengths can only be constructed over countable sets. The core of MML is to quantize the set $\Theta$ to a carefully chosen countable subset $\Theta^* \subset \Theta$ for which a code can be constructed. For each $\bm{\theta}^\prime \in \Theta^*$ there is an associated Voronoi cell $\Omega_{\bm{\theta}^\prime}$, called an \emph{uncertainty region}, containing those distributions $f(\*y | \bm{\theta})$ that are closest in some sense to $f(\*y | \bm{\theta}^\prime)$.
%
%
In the derivation of the MML87 approximation formula, the coding probability $q(\hat{\bm{\theta}})$ of a member $\hat{\bm{\theta}} \in \Theta^*$ is given by
%
%
\begin{equation}
\label{eq:codeprob}
	q(\hat{\bm{\theta}}) \approx \pi(\hat{\bm{\theta}}) w(\hat{\bm{\theta}}) \,,
\end{equation}
where $\pi(\cdot)$ is the Bayesian prior density over the set $\Theta$ and $w(\hat{\bm{\theta}})$ is the volume of the Voronoi cell $\Omega_{\bm{\theta}^\prime}$. In the MML87 approximation, the volume $w(\hat{\bm{\theta}})$ is given by
%
%
\begin{equation}
\label{eq:uncertain}
w(\hat{\bm{\theta}}) = {|\*J(\hat{\bm{\theta}})|}^{-1/2}\kappa_k^{-k/2} \,,
\end{equation} 
where $|\*J(\hat{\bm{\theta}})|$ is the determinant of the Fisher information matrix and $\kappa_k$ is the normalized second moment of an optimal quantizing lattice. The uncertainty region can be interpreted as a set of distributions that are considered indistinguishable on the basis of the observed data and is related to the concept of distinguishable distributions in the MDL literature~\cite{grunwald2005advances}. 

Recall that the MML message consists of two components, the assertion and the detail. From Shannon's theory of information, the length of the assertion, which states the member ${\bm{\theta}} \in \Theta^*$ that will be used to compress the data, is given by $-\log q({\bm{\theta}})$. The length of the detail, which encodes the data using the model named in the assertion, is given by $-\log f(\*y|{\bm{\theta}}) + k/2$ (see~\cite{wallace2005statistical} for more details). Using the MML87 approximation, the codelength of data $\*y$ and a model $\bm{\theta}$ is given by 
\begin{equation}
\label{eq:mml87}
	I_{87}(\*y, \bm{\theta}) = - \log f(\*y|\bm{\theta}) -\log\pi(\bm{\theta}) + \frac{1}{2}\log\abs{\*J_{\bm{\theta}}} \\
	+ \frac{k}{2} \log \kappa_k + \frac{k}{2} \,,
\end{equation}
where $\psi(\cdot)$ is the digamma function and the constant $\kappa_k$ can be approximated by~(\cite{wallace2005statistical}, pp. 257--258)
\begin{equation}
\label{eq:kappa.approx}
\frac{k}{2}(\log\kappa_k+1) \approx -\frac{k}{2}\log(2\pi) + \frac{1}{2}\log(k\pi) + \psi(1) \, .
\end{equation}
%
%
To estimate a model from data using MML, we find the estimate $\hat{\bm{\theta}}$ that minimizes the message length (\ref{eq:mml87}). The key idea behind the MML approach is to minimize the trade-off between the complexity of the model (measured by the assertion length) against the ability of the model to fit the observed data (measured by the detail length). Under suitable regularity conditions, the MML87 codelength is asymptotically equivalent to the well-known Bayesian information criterion~\cite{schwarz1978estimating}, and the MML87 estimator is asymptotically equivalent to the maximum likelihood estimator. 

From (\ref{eq:codeprob}), it is clear that MML uses the prior distribution $\pi(\cdot)$ to construct codewords for members of $\Theta^*$ and is therefore a Bayesian procedure. The MML approach can be seen as performing \emph{maximum a posteriori} (MAP) over the quantized space $\Theta^*$, with the coding probability $q(\bm{\theta})$ playing the role of a discrete prior distribution, and the codelength as a counterpart to the posterior probability. Every model is therefore given a posterior probability mass (i.e., codelength) which allows for discrimination between models with different structures (e.g., non-nested models). A key difference between standard Bayesian estimators such as the MAP and posterior mean estimators, and MML is that MML estimators are parameterization invariant; i.e., they provide the same inferences under one-to-one, continuous re-parameterizations of the parameter space. 



%
\subsection{A small sample message length approximation}
\label{sec:mml:small:sample}
The MML87 approximation relies on a set of regularity conditions on the likelihood function and the prior distributions~\cite{wallace2005statistical}. An important assumption of MML87 is that the prior distribution $\pi(\bm{\theta})$ is approximately uniform over the entire uncertainty region $\Omega_{\bm{\theta}}$. If this condition is violated, the MML87 approximation can break down and the coding probability of the assertion (\ref{eq:codeprob}) can be greater than 1. One solution to this problem, as described in~\cite{wallace2005statistical}, is to use the small-sample message length approximation
\begin{equation}
\label{eq:mmlss}
	I_{\rm{ss}}(\*y,\bm{\theta}) = \frac{1}{2}\log\left(1+\frac{|\*J(\bm{\theta})|\kappa_k^k}{(\pi(\bm{\theta}))^2}\right) - \log f(\*y|\bm{\theta}) + \frac{k}{2} \,,
\end{equation}
which smoothly bounds $q(\bm{\theta})$ to be less than 1. The small sample approximation retains all the attractive properties of the MML87 approximation while being robust to prior distributions that are highly peaked. The cost of introducing additional robustness to the MML87 approximation is that minimizing the small sample codelength (\ref{eq:mmlss}) is, in general, more difficult than minimizing the original MML87 codelength (\ref{eq:mml87}).

%


%
%
%
%
\section{MML Student-\textit{t} linear regression}
\label{sec:linear:t}
%


Given a vector of observations $\*y=(y_1,\dots,y_n)\tran$, a design matrix $\*X_{\gamma} = ( \*x_{\gamma_1},\dots, \*x_{\gamma_p} )$ containing $p = | \gamma |$ predictors indexed by the model structure $\gamma \in \Gamma$ where $\Gamma$ is a set of candidate models, the Student-$t$ linear regression model in (\ref{eq:tlinear}) can be written as:
\begin{equation}
	\label{eq:yi:iid}
	y_i \stackrel{\text{iid}}\sim t(\beta_0 + \bar{\*x}_{i,\gamma} \bbeta, \tau, \nu) \,, \qquad i=1,\dots,n \,.
\end{equation}
where $\bar{\*x}_{i,\gamma} = (x_{i,\gamma_1},\ldots,x_{i,\gamma_p})$ is the $i$-th row of the design matrix $\*X_{\gamma}$. In this paper we assume, without loss of generality, that the columns of the design matrix are standardized to have zero mean. In the remainder of this paper, the model structure index $\gamma$ is omitted when clear from the context. Under the Student-$t$ regression model, the probability density function of $y_i$ is given by
%
\begin{equation}
	f(y_i | \mu_i, \nu, \tau) = \frac{\Gamma\left[(\nu+1)/2\right]}{\Gamma(\nu/2)\sqrt{\pi\nu\tau}} \left[1+\frac{(y_i- \mu_i)^2}{\nu \tau} \right] ^ {-(\nu+1)/2},
\end{equation}
where $\mu_i = \beta_0 + \bar{\*x}_i\tran\bbeta$, $\tau$ is a scale parameter and $\nu$ is the degrees of freedom. The negative log-likelihood of $\*y$, is
\begin{equation}
\label{eq:negloglike}
\begin{aligned}
	-\log f(\*y | \bm{\theta}) & = -n\log\Gamma\left(\frac{\nu+1}{2}\right) + n\log\Gamma\left(\frac{\nu}{2}\right) + \frac{n}{2}\log(\pi\nu\tau)  \\
	  & \quad   +\frac{\nu+1}{2}\sum_{i=1}^{n}\log\left[1+\frac{(y_i-\mu_i)^2}{\nu\tau}\right] \,,
\end{aligned}
\end{equation}
%
%
%
where $\bm{\theta} = (\beta_0, \bm{\beta}, \tau, \nu)$. In order to apply the MML87 approximation (\ref{eq:mml87}), we require the Fisher information matrix for this model. The Fisher information matrix for the Student-$t$ regression model is~\citep{lange1989robust,fonseca2008objective}
\begin{equation*}
	\*J(\bm{\theta}) = 
	\begin{bmatrix}
		\dfrac{n}{\tau}\left(\dfrac{\nu+1}{\nu+3}\right) & \*0 & \*0 \\[2ex]
		\*0 & \dfrac{1}{\tau}\left(\dfrac{\nu+1}{\nu+3}\right)\*X\tran\*X & \*0 \\[2ex]
		\*0 & \*0 & \dfrac{n\nu}{2(\nu+3)\tau^2}
	\end{bmatrix} \,,
\end{equation*} 
and the determinant of the Fisher information matrix is
\begin{equation}
	\label{eq:fish:det}
	\abs{\*J(\bm{\theta})} = \frac{n^2\nu(\nu+1)^{p+1}}{2(\nu+3)^{p+2}\tau^{p+3}}\abs*{\*X\tran\*X} \,.
\end{equation}
Note that the determinant of the Fisher information matrix (\ref{eq:fish:det}) for the Student-$t$ regression model is equal to the determinant of the Fisher information for the Gaussian linear regression model,
\begin{equation}
	\label{eq:fish:det:gauss}
	\abs{\*J_{\textrm{G}}(\bm{\theta})} = \frac{n^2|\*X\tran\*X|}{2\tau^{p+3}} \,,
\end{equation}
scaled by the factor $\nu(\nu+1)^{p+1}/(\nu+3)^{p+2}$ which only depends on $\nu$ and $p$. When the degrees of freedom $\nu \to \infty$, the determinant of the Fisher information matrix (\ref{eq:fish:det}) for the Student-$t$ model reduces to that of the Gaussian model (\ref{eq:fish:det:gauss}).


\subsection{Prior distribution for the regression parameters}
\label{sec:prior}
Recall from Section~\ref{sec:mml} that MML is a Bayesian principle and therefore requires the specification of prior distributions for all model parameters. Assuming that we do not have prior knowledge regarding the values of the coefficients, it is reasonable to give every subset of coefficients $B \subseteq \left\{\beta_1, \dots, \beta_p\right\}$ an equal prior probability. This may be done by placing an improper uniform prior over each coefficient $\beta_i$, i.e.,
\begin{align*}
\pi(\bbeta) \propto 1 \,.
\end{align*}
Using such an improper prior within the MML87 framework results in an infinite message length and bounding of the parameter space is required to ensure propriety. There exist many different approaches to bounding the parameter space; for example, one could use a uniform prior $\beta_i \sim \mathcal{U}(R_a, R_b)$ where $R_b > R_a$. In this manuscript, we propose the following novel hyper-ellipsoid bounding for $\bbeta$:
\begin{equation}
	\label{eq:beta:bound}
	\Lambda(K) = \{\bm{\beta}: \bm{\beta} \in \mathbb{R}^p, \bbeta(\*X\tran\*X)\bbeta\leq K\} \, ,
\end{equation}
%
where $K \in \mathbb R_+$ is a hyperparameter that controls the size of the bounding region; the choice the hyperparameter $K$ is discussed in Section~\ref{ssec:hyperK}. The bounding hyperellipsoid uses intrinsic geometric properties of the model and has an important advantage over the conventional rectangular bounding. Our novel bounding depends on a single hyperparameter which has a natural interpretation: by noting that the quantity $(\*X \bbeta)\tran (\*X\bbeta)$ can be interpreted as the fitted sum-of-squares, the hyperparameter $K$ is seen to directly control the maximum allowable signal strength (i.e., variance).

The set of coefficients that satisfies (\ref{eq:beta:bound}) is called the feasible parameter set and is aligned to the principle axes of the explanatory variables. Using a uniform prior over this feasible parameter set yields the following proper prior distribution for the regression coefficients:
\begin{equation}
	\label{eq:prior:beta}
	\pi(\bbeta | K) = \frac{1}{\text{vol}(\Lambda)} = \frac{\Gamma(p/2+1)\sqrt{\abs{\*X\tran\*X}}}{(\pi K)^{(p/2)}} \,.
\end{equation}
%
For the intercept parameter $\beta_0$, we use the uniform prior 
\begin{equation}
\label{eq:prior:beta0}
	\pi({\beta_0}) \propto 1 \,,
\end{equation}
and for the scale parameter $\tau$, we use the scale-invariant prior distribution
\begin{equation}
	\label{eq:prior:tau}
	\pi(\tau) \propto \frac{1}{\tau} \,.
\end{equation}
The joint prior distribution for all parameters is 
\begin{equation}
	\label{eq:prior:full}
	\pi(\beta_0,\bbeta,\tau | K) = \pi(\beta_0)\pi(\bbeta | K)\pi(\tau) \,.
\end{equation}
The prior distributions for the intercept (\ref{eq:prior:beta0}) and scale parameters (\ref{eq:prior:tau}) are improper and also require a parameter space restriction. However, these parameters are common to all model structures $\gamma \in \Gamma$ and therefore the particular choice of the bounding region does not affect the relative ordering of message lengths. Therefore, the choice of the bounding region for these parameter plays no role in model selection.
%
%

\subsection{Message length for Student-$t$ regression models}
\label{sec:mml.revisit}
The choice of the prior distribution for the regression coefficients $\pi(\bm{\beta} | K)$ is potentially problematic when used in conjunction with the MML87 message length formula. Given a data set that is poorly explained by the predictors (i.e., a fitted signal-to-noise ratio close to zero), the MML87 estimate of the noise variance, $\hat{\tau}$, will be large. From (\ref{eq:uncertain}), we have
\begin{equation}
	w(\hat{\bm{\theta}}) = \left(\frac{\kappa_p\sqrt{2}(\nu+3)^{(p+2)/2}\hat{\tau}^{(p+3)/2}}{n \nu^{1/2}(\nu+1)^{(p+1)/2}}\right) |\*X\tran\*X|  \propto  \hat{\tau}^{(p+3)/2} \,,
\end{equation} 
where $w(\hat{\bm{\theta}})$ is the volume of the uncertainty region used to code the parameter estimates. A large value of $\hat{\tau}$ results in a large volume of the uncertainty region. Recall from Section~\ref{sec:mml} that MML87 approximates the coding probability for the assertion (\ref{eq:codeprob}) by a product of the prior distribution and the volume of the uncertainty region. When the volume of the uncertainty region exceeds the volume of the restricted parameter set $\Lambda(K)$, the coding probability of the assertion will exceed one, which is clearly nonsensical. For any given value of $K$ (i.e., the size of the restricted parameter set), there exists a $\hat{\tau} > 0$ such that $w(\hat{\bm{\theta}}) > {\rm vol}(\Lambda(K))$.


One solution to this problem is to use the small-sample message length approximation formula (\ref{eq:mmlss}) described in Section~\ref{sec:mml:small:sample}. The assertion codelength for $\bm{\beta}$, $I(\bm{\beta}|K)$, is obtained from the small-sample approximation while the assertion codelength for $\beta_0$ and $\tau$, $I(\beta_0, \tau)$, does not depend on $K$ and can be safely obtained using the standard MML87 formula. Given a value of the hyperparameter $K$, the total message length of the data $\*y$ and the parameters $\bm{\theta}$ is
\begin{equation}
	\label{eq:mml.joint}
	I(\*y, \bm{\theta} | K) = I(\bm{\beta}|K) + I(\beta_0, \tau) + I({\bf y} | \bm{\theta}) 
\end{equation}
where
\begin{eqnarray}
	I(\bm{\beta}|K) &=& \frac{1}{2} \log\left\{1+\frac{1}{\left[\Gamma(p/2+1)\right]^2}\left[\frac{\kappa_p\pi K(\nu+1)}{\tau (\nu+3)}\right]^p\right\}  \,, \nonumber \\
	I(\beta_0, \tau) &=& \log(\tau) + \frac{1}{2}\log\left[\frac{n^2\nu(\nu+1)}{2\tau^3(\nu+3)^2}\right] + \frac{1}{2} \log \kappa_2^2 \,, \nonumber  \\
	I(\*y | \bm{\theta}) &=& -n\log\Gamma\left(\frac{\nu+1}{2}\right) + n\log\Gamma\left(\frac{\nu}{2}\right) +\frac{n}{2}\log(\pi\nu\tau) \nonumber \\
	&& +\frac{\nu+1}{2}\sum_{i=1}^{n}\log\left[1+\frac{(y_i-\mu_i)^2}{\nu\tau}\right] + \frac{p+2}{2} \,,
\end{eqnarray}
$\mu_i = \beta_0 + x_i\tran\bbeta$ and the quantization constants can be approximated by
\begin{equation}
\label{eq:kappa.approx}
\kappa_k^k  \approx 2^{-k} k \pi^{1-k} e^{2\psi(1) - k} \, .
\end{equation}
A consequence of using the hyperellipsoidal bounding (\ref{eq:beta:bound}) is that the $|\*X\tran\*X|$ term that appears in both the prior distribution for $\bm{\beta}$ (\ref{eq:prior:beta}) and the Fisher information (\ref{eq:fish:det}) cancels in the formula for the assertion.

\subsection{Estimation of the hyperparameter $K$}
\label{ssec:hyperK}
The message length (\ref{eq:mml.joint}) using the joint prior distribution (\ref{eq:prior:full}) depends crucially on the value of the hyperparameter $K$. For the message to be uniquely decodable, the value of $K$ must be encoded into the assertion component. The complete message length is
\begin{equation}
	\label{eq:complete.msglen}
	I(\*y, \bm{\theta}, K) = I(\*y, \bm{\theta} | K) + I(K)
\end{equation}
where $I(K) = (1/2) \log(n)$ is the codelength for stating the hyperparameter $K$ obtained from the asymptotic codelength approximation for coding continuous parameters~\cite{rissanen1978modeling, schwarz1978estimating}; alternatively, a more accurate codelength could be obtained using the more complex hierarchical message length procedure described in~\cite{makalic2009minimum, schmidt2016approximating}. The MML principle advocates choosing a value for the hyperparameter $K$ which minimizes the message length; i.e., 
\begin{equation}
\label{eq:K:min}
\hat{K} = \argmin_{K > 0}  I(\*y, \hat{\bm{\theta}}(K), K) \, ,
\end{equation}
where $\hat{\bm{\theta}}(K)$ is the MML87 estimate of $\bm{\theta}$ that minimizes (\ref{eq:complete.msglen}) given a value of $K$. As exact minimization of (\ref{eq:K:min}) is a difficult non-convex optimization problem, we use the following approximate procedure. For $K$ sufficiently large, the estimates of $\bm{\beta}$ which minimize the message length are equivalent to the maximum likelihood estimates $\hat{\bm{\beta}}_{\rm ML}$ as neither the prior distribution nor the Fisher information depend on $\bm{\beta}$. An approximate solution to (\ref{eq:K:min}) can be obtaining by requiring that the region $\Lambda(K)$ must include the maximum likelihood estimate $\hat{\bm{\beta}}_{\rm ML}$; this is given by
%
\begin{equation}
\label{eq:K.hat}
\hat{K} = \hat{\bm{\beta}}_{\rm ML}\tran (\*X \tran \*X) \hat{\bm{\beta}}_{\rm ML} \,,
\end{equation}
where $\hat{\bm{\beta}}_{\rm ML}$ is the maximum likelihood estimate of $\bbeta$. 
%
%
%
%
\subsection{MML estimates}
\label{sec:MML:estimates}
The MML estimate $\hat{\bm{\theta}}$ is found by plugging the estimate $\hat{K}$ (\ref{eq:K.hat}) into (\ref{eq:complete.msglen}) and minimizing the resultant message length with respect to $\bm{\theta}$. Dropping terms that do not depend on the parameters $\bm{\theta}$, the message length (\ref{eq:complete.msglen}) is
\begin{align}
\label{eq:msg:onlypara}
I(\*y,\hat{\bm{\theta}},\hat{K}) &= \frac{1}{2}\log\left(1+\frac{B}{\tau^p}\right) + \left(\frac{n-1}{2}\right)\log\tau + \frac{\nu+1}{2}\sum_{i=1}^{n}\log\left[1+\frac{(y_i-\mu_i)^2}{\nu\tau}\right] \,,
\end{align}
where
\begin{equation*}
	B = \frac{1}{\left[\Gamma(p/2+1)\right]^2}\left[\frac{\kappa_p\pi \hat{K}(\nu+1)}{(\nu+3)}\right]^p \,.
\end{equation*}
There is no closed form solution for the the values of $\bm{\theta}$ that minimize (\ref{eq:complete.msglen}). A standard approach to compute the parameter estimates in a Student $t$ regression model is the expectation-maximization (EM) algorithm~\cite{lange1989robust}; other methods include the expectation-conditional maximization algorithm~\cite{liu1995ml} and bootstrapping approaches~\cite{pianto2010bootstrap}. In this paper, we adapt the EM algorithm to find the MML estimates. It is well known that the Student-$t$ distribution can be represented as a scale mixture of normal distributions~\cite{lange1989robust}. We can write model (\ref{eq:yi:iid}) as: 
\begin{eqnarray}
\label{eq:yi:scalenorm}
y_i|\mu_i,\tau,w_i &\sim& \mathcal{N}(\mu_i, \tau/w_i) \,, \\
w_i | \nu &\sim& \text{Gamma}(\nu/2, \nu/2)
\end{eqnarray}
where $w_i$ are the latent (unobserved) variables. Given $y_i$, we have the following property \citep{lange1989robust}:
\begin{equation}
w_i | y_i, \mu_i, \tau, \nu \sim \text{Gamma}(\nu+1,\nu+\delta_i^2) \,,
\end{equation}
where $\delta_i^2 = (y_i-\mu_i)^2/\tau$. The latent variables $w_i$ are not directly observable and will be replaced by the conditional expectation in the E-step of the EM algorithm. The conditional expectation of $w_i$ is
\begin{equation}
\label{eq:em:w.hat}
\hat{w_i} = \EX(w_i|y_i,\bm{\theta}) = \frac{\nu+1}{\nu+\delta_i^2} \,.
\end{equation}
Conditional on the latent variables $w_i$, the message length, up to terms that are independent of $\bm{\theta} = (\beta_0, \bm{\beta}, \tau)$, is given by
%
%
%
%
\begin{equation}
\label{eq:msg.em}
l(\bm{\theta}) = \frac{1}{2}\log\left(1+\frac{B}{\tau^p}\right) + \left(\frac{n-1}{2}\right)\log\tau + \sum_{i=1}^n \frac{w_i(y_i-\mu_i)^2}{2\tau} \,.
\end{equation}
The M-step in the EM algorithm involves solving
\begin{align}
\label{eq:em:beta}
( \hat{\beta}_0, \hat{\bm{\beta}} ) &= \argmin_{{\beta}_0, {\bm{\beta}}} \sum_{i=1}^n \hat{w_i}(y_i-\mu_i)^2 \,, \\
\label{eq:em:tau}
\hat{\tau} &= \argmin_{\tau} l(\tau ; \hat{\beta_0}, \hat{\bbeta}) \,.
\end{align}
MML estimates $\hat{\beta}_0$ and $\hat{\bm{\beta}}$ can be efficiently computed using weighted least squares regression. The function $l(\tau ; \hat{\beta_0}, \hat{\bbeta})$ is log-convex in $\tau$ and the MML estimate may be found by numerical optimization. In summary, the MML estimates are found by iterating the following steps until convergence:
\begin{enumerate}
\item compute the weights $\hat{w}_i$ using (\ref{eq:em:w.hat}); 
\item update $\hat{\beta}_0, \hat{\bm{\beta}}$ using the weights $\hat{w_i}$; 
\item update $\hat{\tau}$ using the current weights $\hat{w_i}$ and estimates $\hat{\beta}_0, \hat{\bm{\beta}}$.
\end{enumerate}
We use the maximum likelihood estimates under the assumption of Gaussian errors as the initial values of $(\hat{\beta}_0, \hat{\bm{\beta}},\hat{\tau})$. 
\subsection{Selection of the degrees of freedom $\nu$}
\label{sec:select_nu}
In order to compute the message length (\ref{eq:complete.msglen}), we require a value for the degrees of freedom parameter $\nu$. A common approach to estimating $\nu$ is to consider only a discrete set $N=\{a_1, \ldots, a_m\}$, where $a_i < a_j$ for all $i<j$, of $m$ candidates values for $\nu \in N$, rather than treating $\nu$ as a continuous parameter. We propose to select the set $N$ such that it includes a range of distributions from those with heavy tails (e.g., the Cauchy distribution when $\nu=1$) to those with light tails (e.g., the normal distribution when $\nu \to \infty$), where each distribution $\nu \in N$ is equidistant from its neighbors. In this paper, the Jensen-Shannon (JS) divergence~\cite{lin1991divergence} is used to measure the distance between two distributions. Let $P$ and $Q$ denote two probability distributions. The Jensen-Shannon divergence from $Q$ to $P$ is given by
\begin{equation}
	\textrm{D}_{\textrm{JS}}(P||Q) = \frac{1}{2}\textrm{D}_{\textrm{KL}}(P||Q) + \frac{1}{2}\textrm{D}_{\textrm{KL}}(Q||P) \,,
\end{equation}
where $\textrm{D}_{\textrm{KL}}(P||Q)$ is the Kullback--Leibler (KL) divergence~\cite{kullback1951information} from $P$ to $Q$, given by
\begin{equation}
	\textrm{D}_{\textrm{KL}}(P||Q) = \int_{-\infty}^{\infty} p(x)\log\frac{p(x)}{q(x)} dx \,,
\end{equation}
where $p(x)$ and $q(x)$ are probability density functions of $P$ and $Q$ respectively. Unlike the KL divergence, the JS divergence is a symmetric function of $P$ and $Q$. Let $t_{\nu}$ denote a standard Student-$t$ probability distribution parameteraised by $\mu=0$, $\tau=1$ and $\nu > 0$ degrees of freedom. We aim to find a candidate set $N$ such that, for all $i \in \{1,\ldots,m-1\}$, 
\begin{equation*}
	\textrm{D}_{\textrm{JS}} (t_{a_i}||t_{a_{i+1}}) = c \, ,
\end{equation*}
where $c>0$ is some constant. In words, this implies that the JS divergence between any two neighbouring distributions $t_{a_i}$ and $t_{a_{i+1}}$ is the same. For example, when $m=4$, the set $N$ that satisfies this condition is $N = \{1, 1.9, 5, \infty \}$. 
%
%
%
%
%
%
%
\subsection{Coding the model index}
\label{sec:model.index}
When deriving the message length (\ref{eq:mml.joint}) in Section~\ref{sec:mml.revisit}, we made the assumption that the model structure $\gamma \in \Gamma$ is known. For the message to be uniquely decodable, the choice of $\gamma$ must be encoded into the assertion component; i.e., 
\begin{equation}
\label{eq:msglen.with.index}
I({\bf y}, \bm{\theta}, K, \gamma) = I({\bf y}, \bm{\theta}, K | \gamma) + I(\gamma) \,,
\end{equation}
where $I(\gamma)$ is the length of the assertion that encodes the model structure $\gamma$. Let $q$ denote the total number of columns (explanatory variables) of ${\bf X}$. We use the following prior distribution over the model structure $\gamma$
\begin{equation}
	\pi(\gamma) = \frac{\pi(|\gamma|)}{Q(|\gamma|)} \, ,
\end{equation}
where $\pi(|\gamma|)$ is a prior distribution over the number of explanatory variables $p=|\gamma|$ (out of $q$) in the model and 
\begin{equation}
	Q(p) = \# \{\gamma \in \Gamma: |\gamma| = p \} \,,
\end{equation}
is the total number of models in $\Gamma$ with $p$ explanatory variables.

In the case nested model selection (e.g., selecting the order of a polynomial), the set $\Gamma = \{\null,\gamma_1,\ldots,\gamma_q\}$ consists of $q$ candidate models such that $\gamma_j = \{1,\ldots,j\}$. Therefore, the total number of models with $p$ explanatory variables is $Q(p) = 1$ for all $p \leq q$. When no prior information regarding the size of the model is available, we choose the uniform prior $\pi(|\gamma|) = 1/(q+1)$ and the assertion length is
\begin{equation}
	I(\gamma) = \log (q + 1) \,.
\end{equation}

In the more general case of all-subsets regression, the set $\Gamma = \mathcal{P}(\left\{1,\dots,q\right\}) \cup \null$ is the set of all subsets of $q$ explanatory variables, including the null model. In this setting, the number of models with $p$ explanatory variables is $Q(p) = {q \choose p}$, and assuming that no prior information is available regarding the size of the model, we again choose the prior distribution $\pi(|\gamma|) = 1/(q+1)$. This yields an assertion of length
\begin{equation}
\label{eq:model.index}
I(\gamma) = \log {q \choose p} + \log (q + 1) \, .
\end{equation}
As discussed in~\cite{scott2010bayes}, this choice of prior distribution implies a marginal prior probability of inclusion of $0.5$ for each explanatory variable. In addition, this prior distribution simultaneously accounts for multiple testing by automatically punishing model sizes that have a large total number of models. 

%
%
%
%
%
\section{Discussion}
\subsection{Behaviour of the dimensionality penalty}
Using the estimate (\ref{eq:K.hat}) of the hyperparameter $K$, we can write 
\begin{equation}
	 \frac{\hat{K}}{\hat{\tau}} = \frac{\hat{\bbeta}\tran(\*X\tran\*X)\hat{\bbeta}}{\hat{\tau}} = n \, \text{SNR} \,,
\end{equation}
where $\hat{\tau}$ is the MML estimate of $\tau$ and SNR is the empirical signal to noise ratio of the fitted Student-\textit{t} regression model. Let 
%
\begin{equation*}
	C_p = \frac{\kappa_p^p\pi^p(\nu+1)^p}{\left[\Gamma(p/2+1)\right]^2(\nu+3)^p n^p} \, .
\end{equation*}
For large \text{SNR}, the codelength of the assertion of the regression parameters $\bm{\beta}$ is
\begin{align}
\label{eq:mmlpen}
I(\bm{\beta} | \hat{K}) =& \frac{1}{2} \log\left\{1 + C_p \left(\frac{\hat{K}}{\hat{\tau}}\right)^p\right\} \nonumber\\
&= \frac{1}{2} \log\left(1+C_p n^p \, {\text{SNR}}^p\right) \nonumber\\
	&\approx  \frac{1}{2}\log C_p + \frac{p}{2} \log(n) + \frac{p}{2}\log(\text{SNR}) \, ,
\end{align}
which, for fixed $n$ and $\tau$, depends only on the empirical SNR. This implies that the better the model fits the data, the larger the penalty for including additional explanatory variables. In contrast, when the empirical \text{SNR} is small (i.e., the model does not fit the data well), the penalty for including an additional explanatory variable is small. This property of the MML criterion is attractive since SNR is a fundamental characteristic of the fitted linear model and is invariant to model parameterization and linear transformations of the data. Therefore, the MML penalty automatically adapts to the amount of detectable signal in the data. The MML regression criterion is not the only model selection technique with this property; the MDL denoising criterion~\citep[p.~117]{rissanen2007information} (see Section \ref{sec:mdl.denoising}) and the recent Bayesian exponentially embedded criterion~\cite{zhu2018bayesian} both exhibit similar behaviour.
\subsection{Gaussian linear regression}
We can specialize the MML criterion for Student-$t$ regression models to the important case of Gaussian linear regression by finding the limit of the message length (\ref{eq:msglen.with.index}) as $\nu \to \infty$; i.e., $I_G({\bf y},\bm{\theta}, K, \gamma) = \lim_{\nu \to \infty} I({\bf y},\bm{\theta}, K, \gamma)$ which is given by 
%
%
\begin{equation}
\label{eq:mml.gauss}
\begin{aligned}
& \frac{n}{2}\log 2\pi\tau + \frac{1}{2\tau}\sum_{i=1}^{n}\left(y_i-\mu_i \right)^2 - \frac{1}{2}\log \tau + \log n + \frac{p}{2} \\
& + \frac{1}{2} \log\left\{1+\frac{\kappa_p^p\pi^p}{\left[\Gamma(p/2+1)\right]^2}\left(\frac{K}{\tau}\right)^p\right\} -\frac{1}{2}\log(4\pi) + I(\gamma)\\
& + \frac{1}{2}\log n + \psi(1) \,,
\end{aligned}
\end{equation}
where $\mu_i = \beta_0 + x_i\tran\bbeta$ and $\psi(\cdot)$ is the digamma function. The hyperparameter $K$ is estimated following the procedure in Section~\ref{ssec:hyperK} and the codelength for the model structure is discussed in Section~\ref{sec:model.index}. 

In comparison to the original MML$_u$ criterion for Gaussian linear regression~\cite{schmidt2009mml}, the Gaussian specialisation of the proposed Student-$t$ criterion (\ref{eq:mml.gauss}) differs primarily in the selection of the hyperparameter $K$ (see Section~\ref{ssec:hyperK}). The original MML$_u$ criterion uses $\hat{K}_u = {\bf y}^\prime {\bf y}$ which for Gaussian models which guarantees that the feasible set $\Lambda_u$ includes the least squares estimates. However, the estimate $\hat{K} = ({\bf X} \hat{\bm{\beta}})^\prime ({\bf X} \hat{\bm{\beta}})$ proposed in this paper results in a feasible set that has a strictly smaller volume than the set $\Lambda_u$ and therefore results in a shorter codelength. As discussed in Section~\ref{sec:mml:small:sample}, the proposed $\hat{K}$ necessitates the use of the small sample MML approximation to avoid problems with the MML87 approximation.
\subsection{Comparison with MDL denoising}
\label{sec:mdl.denoising}
An important consequence of using the hyperellipsoidal prior distribution for the regression parameters (\ref{eq:prior:beta}) is that the MML criterion for Gaussian linear regression (\ref{eq:mml.gauss}) is very similar to the well-known normalized maximum likelihood (NML) regression criterion~\cite{rissanen2000mdl} when the signal to noise ratio is large. In the case of Gaussian linear regression, let 
\begin{equation*}
	C_p = \frac{\kappa_p^p\pi^p}{\left[\Gamma(p/2+1)\right]^2} \, .
\end{equation*}
For large empirical SNR, $\hat{K}/\hat{\tau}$, we have
\begin{equation*}
\begin{aligned}
	\frac{1}{2} \log\left\{ 1 + C_p \left(\frac{\hat{K}}{\tau}\right)^p \right\} &\approx \frac{1}{2} \log C_p  + \frac{p}{2} \log \hat{K} - \frac{p}{2} \log \hat{\tau} \\
	&= \frac{p}{2} \log(\pi\kappa_p \hat{K}) - \frac{p}{2} \log \hat{\tau} \\
	& \quad - \log\Gamma\left(\frac{p}{2}+1\right) \, ,
\end{aligned}
\end{equation*}
and the MML criterion for the Gaussian regression model (\ref{eq:mml.gauss}), ignoring the codelength for the model structure, $I(\gamma)$, can be written as
%
%

%
%
%
\begin{equation}
\label{eq:mml.gauss.simp}
\begin{aligned}
I_G(\*y,\hat{\bm{\theta}}, \hat{K}) \approx & \left(\frac{n-p-1}{2}\right)\left(\log\hat{\tau} + 1\right) + \frac{p}{2}\log(\hat{K})- \log\Gamma\left(\frac{p}{2}+1\right) \\
& + \frac{n}{2}\log 2\pi + \log n + \frac{1}{2}\log p -\frac{p}{2}\log 2 + C_1 \,,
\end{aligned}
\end{equation}
where 
\begin{equation*}
\hat{\tau} = \frac{\sum_{i=1}^{n}\left(y_i-\mu_i \right)^2}{n-p-1} \,,
\end{equation*}
and $C_1$ is a constant that does not depend on the sample size $n$ and dimensionality $p$. This criterion has a similar form to the MDL denoising criterion for the Gaussian linear regression model \citep[p.~117]{rissanen2007information}, which is given by
\begin{equation}
\label{eq:mdl.denoise.temp}
\begin{aligned}
I_{D}(\*y,\bm{\theta},\hat{R})	&= \left(\frac{n-p-1}{2}\right)\log\left(\frac{\hat{\tau}}{n-p-1}\right) + \left(\frac{p+1}{2}\right)\log\frac{\hat{R}}{(p+1)}   \\
	& \quad  + \frac{1}{2}\log[(p+1)(n-p-1)] + \frac{n}{2}\log(2n\pi e) - 3\log 2 + C_2 \,.
\end{aligned}
\end{equation}
%
%
%
where $\hat{R}=n^{-1}\bm{\hat{\beta}}'(\*X'\*X)\bm{\hat{\beta}}=n^{-1}\hat{K}$ and $C_2$ is a constant independent of $n$ and $p$. Using Stirling's approximation, (\ref{eq:mdl.denoise.temp}) can be written as
\begin{equation}
	\label{eq:mdl.denoise.stirling}
	\begin{aligned}
	I_{D}(\*y,\bm{\theta},\hat{K}) &\approx \left(\frac{n-p-1}{2}\right)\log\left(\frac{\hat{\tau}n}{n-p-1}\right) + \left(\frac{p}{2}+\frac{1}{2}\right)\log\hat{K} \\
	& \quad + \frac{n}{2}\log (2\pi) + \frac{1}{2}\log(n-p-1) - \frac{p}{2}\log(p+1) \\
	& \quad + \frac{n}{2} - 3\log 2 + C_2' \,,
	\end{aligned}
\end{equation}
%
where $C_2'$ is another constant that does not depend on $n$ and $p$. Similarly, using Stirling's approximation, the MML criterion (\ref{eq:mml.gauss.simp}) can be written as
\begin{equation}
	\label{eq:mml.gauss.stirling}
	\begin{aligned}
		I_G(\*y,\hat{\bm{\theta}},\hat{K}) &\approx \left(\frac{n-p-1}{2}\right)\log\hat{\tau} + \frac{p}{2}\log\hat{K} + \left(\frac{n-1}{2}\right)\log (2\pi) \\
& \quad  + \log n  + \frac{1}{2}\log p - \left(\frac{p+1}{2}\right)\log(p+2) \\ 
& \quad + \frac{n}{2} + \frac{1}{2}(\log2 + 1) + C_1' \,.
	\end{aligned}
\end{equation}
We can observe that the MML criterion (\ref{eq:mml.gauss.stirling}) is very similar to the MDL criterion (\ref{eq:mdl.denoise.stirling}). Importantly, the fitted signal-to-noise ratio of the model appears in the MDL penalty term. Suppose that $n$ is large and $n \gg p$, the difference between the MML and MDL criteria, for the case of Gaussian noise, is given by
\begin{equation}
	-\frac{1}{2}\log\hat{K} + \frac{1}{2}\left\{\log\left(\frac{np}{p+2}\right) + p\left[\log\left(\frac{p+1}{p+2}\right)\right] \right\} \,.
\end{equation}
When the fitted SNR of the model is small, the penalty term in the MDL criterion is also small, and potentially negative, and this can lead to substantial over-fitting \citep{roos2005behavior}. In contrast, the proposed MML criterion, which makes use of the small sample approximation (\ref{eq:mmlss}), is more robust and does not result in negative codelengths for small SNR (see Section \ref{sec:mml:small:sample} and \ref{sec:mml.revisit}). Furthermore, the MML criterion can be easily extended to other distributions, for example, the Student-$t$ distribution discussed in this paper; such extensions are very difficult in the case of normalized maximum likelihood, due to dependence on the criterion of closed-form expressions for the maximum likelihood estimates. 
\subsection{Model selection consistency}
Recall that, for large empirical SNR, the codelength of the assertion for the regression parameters $\bm{\beta}$ is given by 
\begin{equation}
\label{eq:model.select.consist}
I(\bm{\beta} | \hat{K}) = \frac{1}{2} \log C_p + \frac{p}{2} \log n + \frac{p}{2} \log \left( \frac{\hat{K}}{n \hat{\tau}} \right) \,.
\end{equation}
We make the assumption that the empirical covariance matrix of the predictors satisfies 
\begin{equation}
	\label{eqn:sigma:assumption}
	\lim_{n \to \infty} \left\{\frac{\*X\tran\*X}{n}\right\} = \Sigma \,,
\end{equation}
where $\Sigma$ is a positive definite matrix. As the sample size $n \to \infty$, the estimates $\hat{\beta}$ and $\hat{\tau}$ converge in probability to their true values, and using the assumption (\ref{eqn:sigma:assumption}), it is straightforward to show that 
\begin{align*}
	I(\bm{\beta} | \hat{K}) = \frac{p}{2} \log n + O(1) \, .
\end{align*}
%
%
%
%
For large sample size $n$ and fixed number of predictors $p$, the MML criterion is equivalent to the well-known Bayesian information criterion~\citep{schwarz1978estimating} and is therefore model selection consistent in the sense that as the sample size tends to infinity, the probability of selecting the true model tends to one. However, in contrast to BIC, the proposed MML criterion is also consistent when the sample size $n$ is fixed and the noise variance $\tau \to 0$; i.e., high SNR consistency~\cite{ding2011inconsistency}. This is easily verified by noting that the criterion (\ref{eq:model.select.consist}) satisfies Theorem $1$  in~\cite{schmidt2012consistency}.
%
%
%
%
%
%
%
\subsection{Behaviour under different SNRs}
The MML estimate of $\tau$ exhibits interesting behaviour that depends on the empirical signal strength $\hat{K}$. Recall from Section~\ref{sec:MML:estimates} that the MML estimate of $\tau$ is obtained by minimising (\ref{eq:msg:onlypara}) with respect to $\tau$. For large signal strength $\hat{K}$, (\ref{eq:msg:onlypara}) reduces to
\begin{equation}
l(\bm{\theta}) = \left(\frac{n - p - 1}{2}\right) \log \tau + \frac{1}{2 \tau} \sum_{i=1}^n w_i (y_i - \mu_i)^2 \,,
\end{equation}
and the MML estimate, conditional on $\hat{\bm{\beta}}$ and $\hat{\beta}_0$, is given by 
\begin{equation}
\hat{\tau} = \frac{\sum_{i=1}^n w_i (y_i - \mu_i)^2} {n - p - 1} \, .
\end{equation}
In this setting, the MML estimate $\hat{\tau}$ is equivalent to the maximum likelihood estimate of $\tau$ adjusted to take into account that we are fitting $p$ predictors (and the intercept parameter) to the data. In contrast, when the signal strength $\hat{K}$ is close to zero, (\ref{eq:msg:onlypara}) simplifies and the MML estimate reduces to
%
%
%
\begin{equation}
\hat{\tau} = \frac{\sum_{i=1}^n w_i (y_i - \mu_i)^2} {n - 1} \, ,
\end{equation}
which is equivalent to the maximum likelihood estimate of $\tau$ adjusted only for the intercept parameter. The MML estimate depends on the signal strength and automatically varies the degrees of freedom used to adjust the maximum likelihood estimate; when there is little signal, predictors are not used to fit the data and thus there is no need to adjust the maximum likelihood estimate. Alternatively, when there is strong signal, the MML estimate appropriately adjusts for all $p$ degrees of freedom, with a smooth transition between these two regimes.

\section{Results}
\label{sec:results}
\subsection{Simulation}
To evaluate the performance of the proposed MML criterion we performed several simulation studies. We compared our new criterion against the Bayesian information criterion (BIC)~\citep{schwarz1978estimating} and the Akaike information criterion with a correction for finite sample sizes (AIC$_c$)~\citep{hurvich1993corrected}, given by
\begin{align*}
	\mathrm{BIC} (\*y,\hat{\bm{\theta}}_{\rm ML})  &= L(\*y,\hat{\bm{\theta}}_{\rm ML}) + \frac{k}{2} \log n \,, \\
	\mathrm{AIC}_c(\*y,\hat{\bm{\theta}}_{\rm ML}) &= L(\*y,\hat{\bm{\theta}}_{\rm ML}) + k + \frac{2k(k+1)}{n-k-1} \,,
\end{align*}
where $L(\cdot)$ is the negative log-likelihood of the Student-\textit{t} regression model, which is given by (\ref{eq:negloglike}), $\hat{\bm{\theta}}_{\rm ML}$ is the maximum likelihood estimate of $(\beta_0,\bm{\beta},\tau,\nu)$ and $k = p + 2$ is the number of free parameters. 

The simulation was performed as follows. We generated training samples with sample size $n=50$ from a Student-\textit{t} regression model with varying degrees of freedom $\nu$.  Model selection was performed using the aforementioned model selection criteria: (i) MML, (ii) BIC, and (iii) AIC$_c$. The degrees of freedom were estimated from the set $\nu \in \{1, 1.9, 5, \infty \}$; this set includes a range of distributions from those with very heavy tails through to the Gaussian distribution (see ~Section \ref{sec:select_nu}). For each experiment, we computed the lasso regression path~\cite{tibshirani96regression}. For each model structure in the path, the model coefficients were estimated using both MML and maximum likelihood. In the case of the MML criterion, the model structure was encoded using the nested prior specification (see Section~\ref{sec:model.index}). Each of the model selection criteria was then used to nominate a best model (i.e., the set of predictors and the degrees of freedom that yielded the lowest criterion score) from this path.

For each experiment, the design matrix $\*X$ for both the training and testing samples were generated by a multivariate normal distribution with mean $\bm{0}$ and covariance matrix $Q$, where $Q_{ij}=0.5^{|i-j|}$. The sample size of the training and testing sample were $50$ and $10000$ respectively. The dimensionality of $\bm{\beta}$ was chosen to be $15$. The number of non-zero entries of $\bm{\beta}$ was chosen to be $15$ (dense model), $8$ (balanced model) and $3$ (sparse model). For each degree of sparsity, three different signal strengths were tested: (1)~$\beta_j=2$ (strong), (2)~$\beta_j=1$ (moderate), and (3)~$\beta_j=0.5$ (weak). For each choice of sparsity and signal strength, the degrees of freedom of the generating model were selected from the set $\{1,5,\infty\}$. Each simulation experiment (combination of sparsity, signal strength and true degrees of freedom) was repeated $100$ times. 

\begin{table}[htbp]
\setlength\tabcolsep{3.5pt} 
\footnotesize
\centering
\caption{Average empirical KL divergence, absolute prediction errors and average number of false positives and false negatives for MML, BIC and AIC$_c$ in $100$ experiments. In each experiment, the training sample size was $n=50$.}
\label{tab:sim}

\sisetup{
  table-align-uncertainty=true,
  separate-uncertainty=true,
}
\renewrobustcmd{\bfseries}{\fontseries{b}\selectfont}
\renewrobustcmd{\boldmath}{}

\hspace*{-0.4cm}\begin{tabular}{lllccc@{\hspace{0.65cm}}ccc@{\hspace{0.65cm}}ccc@{\hspace{0.65cm}}ccc}
\toprule
	\addlinespace[0.2em]
	\multicolumn{3}{c}{True model}
	& \multicolumn{3}{c}{KL divergence}
	& \multicolumn{3}{c}{Prediction errors} 
	& \multicolumn{3}{c}{False positive}
	& \multicolumn{3}{c}{False negative} \\
	d.f. & Sparsity & Signal & MML & BIC & AIC$_c$ & MML & BIC & AIC$_c$ & MML & BIC & AIC$_c$ & MML & BIC & AIC$_c$ \\ 
	\midrule
	\multirow{9}{*}[-6pt]{t(1)} & \multirow{3}{*}{Dense} & Weak & \bfseries 0.367 & 0.857 & 0.713 & \bfseries 10.80 & 10.86 & 10.90 & 0 & 0 & 0 & 4.74 & \bfseries 2.90 & 5.52 \\
	 & & Moderate  & \bfseries 0.458 & 0.906 & 0.905 & 10.72 & \bfseries 10.56 & 10.80 & 0 & 0 & 0 & 3.09 & \bfseries 0.80 & 2.70 \\
	 & & Strong   & \bfseries 0.385 & 0.922 & 0.932 & \bfseries 13.15 & 13.16 & 13.20 & 0 & 0 & 0 & 0.63 & \bfseries 0.30 & 0.42 \\
	 \addlinespace[0.6em]
	 & \multirow{3}{*}{Balanced} & Weak & \bfseries 0.292 & 0.586 & 0.435 & \bfseries 21.65 & 21.74 & 21.72 & 3.52 & 3.13 & \bfseries 2.04 & \bfseries 2.81 & 3.47 & 4.51 \\
	 & & Moderate  & \bfseries 0.381 & 0.799 & 0.660 & \bfseries 14.38 & 14.44 & 14.48 & 3.40 & 4.65 & \bfseries 3.17 & 1.84 & \bfseries 1.22 & 2.08 \\
	 & & Strong   & \bfseries 0.320 & 0.732 & 0.659 & 12.84 & \bfseries 12.82 & 12.88 & \bfseries 3.72 & 5.14 & 4.07 & 0.51 & \bfseries 0.15 & 0.37 \\
	 \addlinespace[0.6em]
	 & \multirow{3}{*}{Sparse} & Weak & \bfseries 0.208 & 0.301 & 0.248 & 11.97 & 11.97 & \bfseries 11.94 & 4.86 & 2.18 & \bfseries 1.92 & \bfseries 1.20 & 2.02 & 1.98 \\
	 & & Moderate  & \bfseries 0.272 & 0.485 & 0.383 & \bfseries 9.619 & 9.676 & 9.678 & 5.75 & 4.56 & \bfseries 3.24 & \bfseries 0.54 & 0.92 & 1.21 \\
	 & & Strong   & \bfseries 0.207 & 0.430 & 0.358 & \bfseries 8.379 & 8.435 & 8.419 & \bfseries 4.52 & 5.55 & 4.64 & 0.15 & \bfseries 0.06 & 0.16  \\
	 \midrule
	 \addlinespace[0.5em]
	\multirow{9}{*}[-6pt]{t(5)} & \multirow{3}{*}{Dense} & Weak & \bfseries 0.323 & 1.105 & 1.133 & \bfseries 1.216 & 1.347 & 1.387 & 0 & 0 & 0 & 4.28 & \bfseries 2.74 & 4.27 \\
	 & & Moderate  & \bfseries 0.287 & 1.040 & 1.057 & \bfseries 1.168 & 1.306 & 1.319 & 0 & 0 & 0 & 0.57 & \bfseries 0.26 & 0.47 \\
	 & & Strong   & \bfseries 0.271 & 1.085 & 1.085 & \bfseries 1.181 & 1.317 & 1.317 & 0 & 0 & 0 & \bfseries 0 & \bfseries 0 & 0.03 \\
	 \addlinespace[0.6em]
	 & \multirow{3}{*}{Balanced} & Weak & \bfseries 0.282 & 0.492 & 0.405 & \bfseries 1.186 & 1.231 & 1.224 & 1.17 & 1.59 & \bfseries 1.08 & \bfseries 1.85 & 1.87 & 2.33 \\
	 & & Moderate  & \bfseries 0.241 & 0.562 & 0.419 & \bfseries 1.130 & 1.178 & 1.152 & \bfseries 1.01 & 2.22 & 1.33 & 0.20 & \bfseries 0.09 & 0.19 \\
	 & & Strong   & \bfseries 0.172 & 0.477 & 0.327 & \bfseries 1.075 & 1.150 & 1.108 & \bfseries 0.61 & 1.72 & 0.92 & \bfseries 0 & \bfseries 0 & \bfseries 0 \\
	 \addlinespace[0.6em]
	 & \multirow{3}{*}{Sparse} & Weak & \bfseries 0.196 & 0.254 & 0.229 & \bfseries 1.076 & 1.088 & 1.078 & 1.83 & 1.19 & \bfseries 1.17 & \bfseries 0.53 & 0.90 & 0.77 \\
	 & & Moderate  & \bfseries 0.131 & 0.233 & 0.176 & \bfseries 1.030 & 1.049 & 1.037 & \bfseries 0.69 & 1.24 & 0.89 & \bfseries 0.01 & \bfseries 0.01 & \bfseries 0.01 \\
	 & & Strong   & \bfseries 0.094 & 0.136 & 0.142 & \bfseries 1.003 & 1.017 & 1.022 & \bfseries 0.17 & 0.54 & 0.61 & \bfseries 0 & \bfseries 0 & \bfseries 0 \\
	 	 \midrule
	 \addlinespace[0.5em]
	\multirow{9}{*}[-6pt]{t($\infty$)} & \multirow{3}{*}{Dense} & Weak & \bfseries 0.443 & 1.060 & 0.984 & \bfseries 1.106 & 1.132 & 1.195 & 0 & 0 & 0 & 2.95 & \bfseries 1.45 & 3.07 \\
	 & & Moderate  & \bfseries 0.266 & 1.086 & 1.146 & \bfseries 0.978 & 1.113 & 1.146 & 0 & 0 & 0 & \bfseries 0.09 & \bfseries 0.09 & 0.23 \\
	 & & Strong   & \bfseries 0.295 & 1.080 & 1.080 & \bfseries 0.987 & 1.107 & 1.107 & 0 & 0 & 0 & \bfseries 0 & \bfseries 0 & \bfseries 0 \\
	 \addlinespace[0.6em]
	 & \multirow{3}{*}{Balanced} & Weak & \bfseries 0.290 & 0.491 & 0.414 & \bfseries 0.980 & 1.010 & 1.005 & 1.26 & 1.69 & \bfseries 1.14 & 1.38 & \bfseries 1.26 & 1.6 \\
	 & & Moderate  & \bfseries 0.228 & 0.506 & 0.390 & \bfseries 0.927 & 0.979 & 0.950 & \bfseries 1.07 & 1.87 & 1.36 & 0.08 & \bfseries 0.03 & 0.05 \\
	 & & Strong   & \bfseries 0.152 & 0.444 & 0.285 & \bfseries 0.890 & 0.956 & 0.917 & \bfseries 0.47 & 1.63 & 0.75 & \bfseries 0 & \bfseries 0 & \bfseries 0\\
	 \addlinespace[0.6em]
	 & \multirow{3}{*}{Sparse} & Weak & \bfseries 0.172 & 0.193 & 0.179 & 0.901 & 0.903 & \bfseries 0.900 & 1.62 & \bfseries 1.10 & 1.17 & \bfseries 0.39 & 0.58 & 0.50 \\
	 & & Moderate  & \bfseries 0.114 & 0.161 & 0.163 & \bfseries 0.860 & 0.870 & 0.872 & \bfseries 0.70 & 0.97 & 1.05 & \bfseries 0 & \bfseries 0 & \bfseries 0 \\
	 & & Strong   & \bfseries 0.097 & 0.164 & 0.149 & \bfseries 0.848 & 0.867 & 0.866 & \bfseries 0.32 & 0.87 & 0.81 & \bfseries 0 & \bfseries 0 & \bfseries 0 \\
\bottomrule
\end{tabular}
\end{table}

\begin{table}[H]
\centering
\caption{Posterior marginal inclusion probabilities of the $13$ predictors in the Boston housing data.}
\label{tab:boston_rank}
\resizebox{\textwidth}{!} {%
\begin{tabular}{cccccccccccccc}
\toprule
& INDUS & CHAS & AGE& ZN & CRIM & NOX & RAD & TAX & BLACK & DIS & LSTAT & PTRATIO & RM  \\  
\midrule
Marginal Probability & 0.308 & 0.804 & 0.900 & 0.990 & 0.998 & 0.998 & 0.999 & 0.999 & 1.000 & 1.000 & 1.000 & 1.000 & 1.000 \\
\bottomrule
\end{tabular}%
}
\end{table}
The performance of each criterion was measured by: (i) the Kullback--Leibler (KL) divergence \citep{kullback1951information} from the selected model to the true generating model, (ii)~the mean absolute prediction error, (iii)~the number of predictors erroneously included in the model (false positives), and (iv)~the number of associated predictors erroneously excluded from the model (false negatives). The results are presented in Table~\ref{tab:sim}. 

In terms of KL divergence, MML performed substantially better than BIC and AIC$_c$ in all tests and obtained particularly good performance when the true generating model was dense. In general, MML also achieved the best performance in terms of prediction error, particularly when the signal strength is weak. In contrast, it is not clear which criterion has the overall best performance in terms of the number of false positives and false negatives. In general, the number of false positives decreases when the sparsity of the model increases for all three criteria, except for the case when the true degrees of freedom is $\nu = 1$. In this case, MML tends to overfit more than the other two criteria when the signal strength is weak or moderate. As discussed in Section~\ref{sec:mml.revisit}, when the signal is weak, the MML criterion assigns a smaller penalty to each predictor in the model and ``false'' predictors are therefore more likely to be included in the best model. Although this leads to some overfitting, it has little negative impact on prediction as the signal being fitted is weak. In contrast, when the signal is strong, the MML criterion generally obtained the smallest number of false positives. The universally strong predictive performance of MML highlights the difference in aims between selecting the true model versus making good predictions, in the sense that even in the case that BIC and AIC$_c$ achieved less false positives and negatives, they usually performed worse in terms of both prediction metrics. 

\subsection{Real data}
We applied our proposed MML model section criterion to the Boston housing data. This data reports the prices of owner-occupied homes in the Boston area. It contains $n=506$ samples and $q=13$ predictors. The response variable is the median value of owner-occupied homes in $\$1,000$s. The degrees of freedom $\nu$ was estimated from the set $\nu \in \{1, 1.9, 5, \infty \}$. We performed an exhaustive search over the space of $8,192$ model structures (i.e., all subset selection) and computed the message length for each model. The model with the shortest message length ($1,468$ nits) included $12$ predictors with degrees of freedom $\nu = 1.9$. In contrast, the best Gaussian model had a message length of $1,549$ nits, which is significantly longer; this strongly suggests that a heavy-tailed error model is appropriate for this data.

A particular advantage of MML is that codelengths can be interpreted as posterior probabilities and this can be used to create a posterior distribution over the space of $8,192$ model structures. Let $\mathcal{M}=\{M_1,\dots,M_{8192}\}$ be the model space and $I(M)$ be the minimum message length achievable by model $M$. The posterior probability assigned to model $M_i$ by MML is given by
\begin{equation}
	P(M_i) = \frac{\exp(-I(M_i))}{\sum_{M \in \mathcal{M}} \exp(-I(M))} \,. 
\end{equation}
Using these posterior probabilities, we computed the marginal probability of inclusion for each of the $13$ predictors (see Table~\ref{tab:boston_rank}). We can observe that $12$ out of the $13$ predictors have marginal probability greater than $0.5$. Only the predictor recording the proportion of non-retail business acres per town ({\tt indus}), which had a marginal probability of $0.308$, was deemed unassociated.

To test the predictive performance of the proposed MML criterion, we performed another set of simulations. The Boston housing data was divided randomly into training and testing sets of equal size. We performed model selection using the training samples and tested the predictive performance of the MML, BIC and AIC$_c$ criteria on the testing samples. All three criteria selected $\nu=1.9$ for $90\%$ of the tests and never selected the Gaussian model. The prediction performance is summarised in Table~\ref{tab:boston_predict}. The models selected by MML attained, on average, a smaller negative log-likelihood for the testing data while AIC$_c$ marginally outperformed MML in terms of absolute prediction error.

%

\begin{table}[t]
\centering
\small
\caption{Average negative log-likelihood (NLL) and absolute prediction error for MML, BIC and AIC$_c$ in the Boston housing data, obtained from 50 cross-validation experiments. The training and testing sample were both of size $n=253$.}
\label{tab:boston_predict}
\begin{tabular}{llcccc}
\toprule
 & & t(1) & t(1.9) & t(5) & t($\infty$) \\
\midrule
\multirow{2}{*}{MML} & NLL & 2.9471 & 2.8892 & 2.9080 & 3.0494 \\ 
& Pred. Error & 3.3736 & 3.3572 & 3.3449 & 3.4869 \\
\addlinespace[0.8em]
\multirow{2}{*}{BIC} & NLL & 2.9696 & 2.9096 & 2.9329 & 3.0723 \\ 
& Pred. Error & 3.4261 & 3.4082 & 3.4142 & 3.5571 \\
\addlinespace[0.8em]
\multirow{2}{*}{AICc} & NLL & 2.9534 & 2.8894 & 2.9096 & 3.0484 \\ 
& Pred. Error & 3.3697 & 3.3388 & 3.3382 & 3.4777 \\
\bottomrule
\end{tabular}
\end{table}
%
%
%
%
%
%
\afterpage{\clearpage}
\bibliographystyle{IEEEtranN}
\bibliography{mybibfile}
\end{document}